\documentclass[submission,copyright,creativecommons]{eptcs}
\providecommand{\event}{TAV-WEB 2010} 
\usepackage{breakurl}             

\usepackage{amsmath}
\usepackage{amssymb}
\usepackage{amsfonts}
\usepackage{bm}

\usepackage{multicol}

\usepackage{bcprules}
\usepackage{listings}

\usepackage{graphicx}
\usepackage{subfigure}

\usepackage{amsthm}

\title{Analysis and Verification of Service Interaction Protocols \\ -- A
  Brief Survey --}

\author{
Gwen Sala\"un
\institute{Grenoble INP--INRIA--LIG, France}
\email{Gwen.Salaun@inria.fr}
}
\def\titlerunning{Analysis and Verification of Services}
\def\authorrunning{Gwen Sala\"un}
\begin{document}
\maketitle

\begin{abstract}
  Modeling and analysis of interactions among services is a crucial
  issue in Service-Oriented Computing. Composing Web services is a
  complicated task which requires techniques and tools to verify that
  the new system will behave correctly.  In this paper, we first
  overview some formal models proposed in the literature to describe
  services. Second, we give a brief survey of verification techniques
  that can be used to analyse services and their interaction.  Last, we
  focus on the realizability and conformance of choreographies.
\end{abstract}

\section{Introduction}
\label{sec:introduction}

Service-Oriented Computing (SOC) has emerged as a new software
development paradigm that enables implementation of Web accessible
software systems that are composed of distributed services which
interact with each other via the exchange of messages.  In order to
facilitate integration of independently developed services that may
reside in different organizations, it is necessary to provide some
analysis and verification techniques to check as automatically as
possible that the new system will behave correctly avoiding erroneous
interactions leading to deadlock states for instance.

\begin{figure}
\centering
\includegraphics[width=0.6\linewidth,clip]{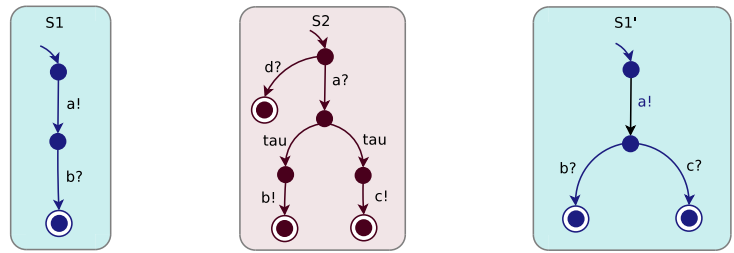}
\caption{Deadlocking execution of services}
\label{fig:unspecified}
\end{figure}

\begin{figure}
\centering
\includegraphics[width=0.6\linewidth,clip]{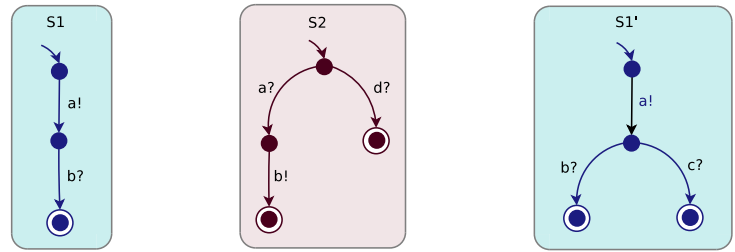}
\caption{Unmatching messages}
\label{fig:sim}
\end{figure}

Let us show a couple of examples to illustrate the previous arguments,
where services are modelled using Labelled Transition Systems
(presented more formally in Section~\ref{section:models}).  Services
{\sf S1} and {\sf S2} in Figure~\ref{fig:unspecified} can end up into
a deadlock because after interacting on {\sf a}, {\sf S2} can decide
to evolve through an internal action $\tau$ (right-hand branch of the
choice) and is deadlocked: {\sf S1} cannot interact on {\sf c} with
{\sf S2} at this point. On the other hand, the execution of {\sf S1'}
and {\sf S2} is free of deadlocks because all emissions on both sides
have a matching reception on the other.
In Figure~\ref{fig:sim}, suppose that {\sf S1} is a client and {\sf
  S2} a service. {\sf S1} is satisfied because the service is able to
reply his/her request, {\it i.e.}, can receive {\sf a} and send {\sf
  b}. However, if we focus on another version of this client {\sf
  S1'}, after submitting {\sf a}, the client expects either {\sf b} or
{\sf c}, but {\sf S2} is not able to provide {\sf c}. This is another
kind of issue that one may need to detect: all the messages (in the
client here) must have a counterpart.

In this paper, we do not want to present the many works and papers
which have proposed analysis and verification for Web services, this
would be too long and uninteresting. Our goal is to focus on specific
issues occuring in this area, and present some automated techniques to
work them out. We will also give some key references for each problem
to enable the reader to go deeper in these issues and solutions
existing for them. Beyond giving a quick overview of service analysis
techniques, we also point out at the end of the paper a few challenges
that are still open, to the best of our knowledge.

The organization of this paper is as follows.  First, we present in
Section~\ref{section:models} some formal models that are often used to
represent abstract descriptions of services, {\it e.g.}, Petri nets,
automata-based models, process algebras. In
Section~\ref{section:verification}, we focus on automated verification
techniques, namely equivalence-checking on one hand, and temporal
properties and model-checking on the other.
Section~\ref{section:compatibility} is dedicated to the compatibility
of two (or more) services. This section also comments on some
techniques to quantify the compatibility degree between two services,
and on service adaptation which is a solution to work out existing
mismatches detected using compatibility analysis. In
Section~\ref{section:realizability}, we present a slightly different
kind of analysis which aims at checking the realizability (and
conformance) of choreography specifications. Realizability indicates
whether services can be generated from a given choreography
specification in such a way that the interactions of these services
exactly match the choreography specification. Finally, we draw up some
conclusions in Section~\ref{section:conclusion}.

\section{Models of Services}
\label{section:models}

In this section, we focus on formal models. Bringing formality to the
service development process opens the way to the writing and
verification of properties that the designer expects from his/her
system. This is not the case of semi-formal notations such as UML or
BPMN which are often acknowledged as more readable and user-friendly
than formal methods but lack formal semantics and validation tools.
Services are distributed components which communicate exchanging
messages, therefore they are best described using behavioural
description languages. Several candidates have been used in the
literature:

\begin{itemize}
  \item Process algebras (or calculi): CCS, CSP, LOTOS, FSP, etc
  \item Automata-based models: state diagrams,
  Harel's Statecharts, IO-Automata, LTS, etc
  \item Petri nets: coloured Petri nets, workflow nets, open nets, etc
  \item Temporal Logic: Lamport's TLA
  \item Message Sequence Charts
\end{itemize}

Here are a few
references~\cite{BultanWWW04,SalaunBS06,LTSAWS,vanBreugel2006,AalstMSW09}
where the reader can find more details about these models and their
use in the service development process. According to us, process
algebras are one of the best candidates to specify service models for
four reasons: (i)~the existing calculi present several levels of
abstraction useful to have a more faithful representation of a
service, {\it e.g.}, specifying data (LOTOS) or mobility
($\pi$-calculus), (ii)~they are compositional notations, then adequate
to describe composition of services, (iii)~they provide textual
notation which makes them scalable to tackle real-world systems, and
(iv)~there exist some state-of-the-art verification tool-boxes for
these languages, {\it e.g.}, SPIN, CADP, UPPAAL, or $\mu$-CRL2.

In the rest of this paper, for illustration purposes and for the sake
of readability (process algebras are not perfect, unfortunately), we
assume that services are modelled using {\it Labelled Transition
  Systems} (LTSs). An LTS is a tuple $(A, S, I, F, T)$ where: $A$ is
an alphabet which corresponds to the set of labels associated to
transitions, $S$ is a set of states, $I \in S$ is the initial state,
$F \subseteq S$ is a nonempty set of final states, and $T \subseteq S
\times A \times S$ is the transition relation.  In our model, a {\it
  label} is either a $\tau$ (internal action) or a tuple $(m,d)$ where
$m$ is the message name, and $d$ stands for the communication
direction (either an emission $!$ or a reception $?$).  Labels can
take typed parameters or arguments into account as well, and in such a
case the transition system is called {\it symbolic} (STS).  Using this
model, a choice can be represented using either a state and at least
two outgoing transitions labelled with observable actions (external
choice) or branches of $\tau$ transitions (internal choice). LTSs and
STSs can be easily derived from higher-level description languages
such as Abstract BPEL, see for
instance~\cite{BultanWWW04,SalaunBS06,CamaraMSCOCP09} where such
abstractions were used for verification, composition or adaptation of
Web services. The operational semantics of STSs is given
in~\cite{DOS-Foclasa09}.

Several communication models can be assumed among services. In
particular, we would like to say a word here about synchronous {\it
  vs.} asynchronous communication. Synchronous communication
corresponds to handshake communication whereas asynchronous
communication uses message queues for interaction purposes (similarly
to mailboxes). Most existing works rely on synchronous communication.
Asynchronous communication is as realistic as synchronous
communication, however, results are more complicated to obtain and
even sometimes undecidable~\cite{DZ-ACM83} (see
Section~\ref{section:conclusion} for a more detailed discussion). In
this paper, we assume a binary communication model where two services
synchronize if one can evolve through an emission, the other through a
reception, and both labels share the same message.

\medskip

\begin{figure}[!ht]
\centering
\includegraphics[width=0.7\linewidth,clip]{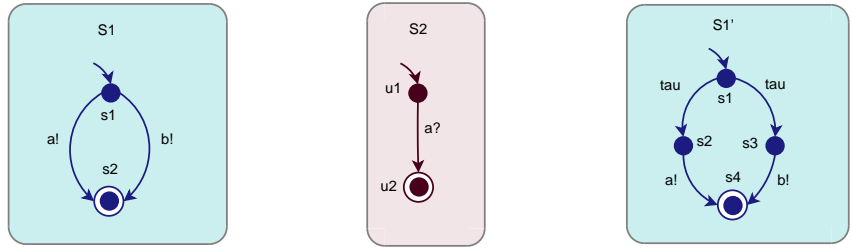}
\caption{{\sf S1} and {\sf S2} interoperate successfully, but {\sf S1'} and {\sf S2} can deadlock}
\label{fig:internal-vs-external}
\end{figure}

{\bf Internal behaviours.}  Service analysis could be worked out
without taking into account their internal evolution because that
information is not observable from its partners point of view
(black-box assumption). However, keeping an abstract description of
the non-observable behaviours while analysing services helps to find
out possible interoperability issues.  Indeed, although one service
can behave as expected by its partner from an external point of view,
interoperability issues may occur because of unexpected internal
behaviours that services can execute.  For instance,
Figure~\ref{fig:internal-vs-external} shows two versions of one
service protocol without ({\sf S1}) and with ({\sf S1'}) its internal
behaviour. Assuming a synchronous communication model, {\sf S1} and
{\sf S2} can interoperate on {\sf a} and terminate in final states
({\sf b!} in {\sf S1} has no counterpart in {\sf S2} and cannot be
executed). However, if we consider {\sf S1'}, which is an abstraction
closer to what the service actually does, we see that this protocol
can (choose to) execute a $\tau$ transition at state {\sf s1} and
arrives at state {\sf s3} while {\sf S2} is still in state {\sf u1}.
At this point, both {\sf S1'} and {\sf S2} cannot exchange messages,
and the system deadlocks. This issue would not have been detected with
{\sf S1}.

The reader interested in more details about $\tau$ transitions and
their handling can refer to~\cite{OS-WCSI10}.

\section{Automated Verification}
\label{section:verification}

A major interest of using abstract languages grounded on a clear
semantics is that automated tools can be used to check that a system
matches its requirements and operates safely. Specifically, these
tools can help (i)~checking that two services are in some precise
sense \emph{equivalent} -- one behaviour is typically a very abstract
one expressing the specification of the problem, while the other is
closer to the implementation level; this can also be used for checking
the substitutability (or replaceability) of one service by another;
(ii)~checking that a service (possibly composite) verifies desirable
\emph{properties} -- {\it e.g.}, the property that the system will
never reach some unexpected state. Revealing that the composition of a
number of existing services does not match an abstract specification
of what is desired, or that it violates a property which is absolutely
needed can be helpful to correct a design or to diagnose bugs in an
existing service. Note that in the following of this section, we focus
on verification techniques that are of interest for Web services, and
we do not give an overview of the many papers that have been published
on this topic (most of them do the same using different languages and
tools), see for
instance~\cite{BultanWWW04,SalaunBS06,LTSAWS,vanBreugel2006,CDRM-JLAP-2010}.

\subsection{Verifying Equivalences}

Intuitively, two services are considered to be equivalent
if they are \emph{indistinguishable} from the viewpoint of an external
observer interacting with them. This notion has been formally defined
in the process algebra community, and several notions of equivalence
have been proposed~\cite{MilCC}. Equivalences are strong yet suitable
relations for these checks, because they preserve all observable
actions. However, these notions exhibit some subtleties relevant to
the context of Web services.

\begin{figure}[!ht]
\centering
\includegraphics[width=0.9\linewidth,clip]{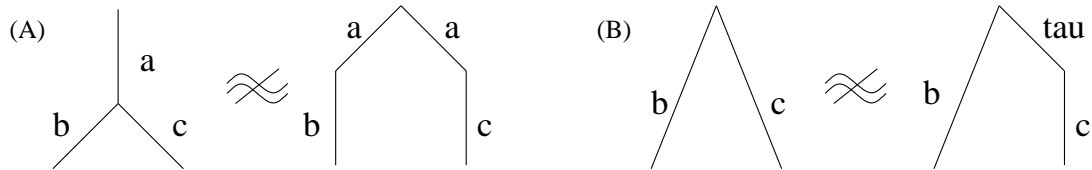}
  \caption{Classical examples of services not observationally equivalent.}
  \label{figure:traces}
\end{figure}

A first approach is to consider two services to be equivalent if the
set of \emph{traces} they can produce is the same
(\emph{trace-equivalence}). For instance, the possible executions of
the services shown in Fig.~\ref{figure:traces} part (A), where
messages {\sf a}, {\sf b} and {\sf c} can be respectively understood
as requests for reservation, editing data and cancellation. Both of
these two services will have {\sf a.b} and {\sf a.c} as possible
traces: they will either receive the messages {\sf a} then {\sf b}, or
{\sf a} then {\sf c}.

Nevertheless, it is not fully satisfactory to consider these two
services equivalent since they exhibit the following subtle
difference. After receiving message {\sf a}, the first service will
accept either message {\sf b} or {\sf c}. The second service behaves
differently: on receiving message {\sf a}, it will either choose to
move to a state where it expects message {\sf b}, or to a state where
it expects message {\sf c}. Depending on the choice it makes, it will
not accept one of the messages whereas the first service leaves both
possibilities open. The second service does not guarantee that a
request for reservation ({\sf a}) followed by, {\it e.g.},
cancellation ({\sf c}) will be handled correctly ({\sf c} might not be
possible if the service has chosen the left-hand side branch).  The
notion of equivalence called \emph{bisimulation}~\cite{MilCC} is a
refinement of trace equivalence which takes these differences into
account.

Further subtleties arise when one has a partial knowledge of the
service behaviour. This may happen for two reasons: (i)~during the
design stage, where the specification which is being defined is
abstract and incomplete; (ii)~when one finds or reuses an existing
service, and only an interface or a partial description hiding private
details is available. $\tau$ actions must be taken into account when
reasoning on the equivalence of two services, as evidenced by
Fig.~\ref{figure:traces} part (B). Both of the services depicted here
can receive {\sf b} (edition of reservation data) or {\sf c}
(cancellation). But whereas the first one can receive any of the two,
the second one can choose to first execute some unobservable action
which will lead it to a state where it can only receive message {\sf
  c}.  Once again it cannot be guaranteed that the second service will
accept cancellation requests, and this depends on some decisions it
takes internally.

Weak (or observational) and branching equivalences are the strongest
of the weak equivalences~\cite{ChapterHPA-Intro}, branching
equivalence being the strongest of these two. They preserve
behavioural properties (do not add deadlocks for instance) on
observable actions, and are therefore acknowledged as the most
appropriate notions of process equivalence, in the context of Web
services. They are implemented in tools like CADP~\cite{CADP2006}
which can automatically check that two transition systems denote the
same observational (or branching) behaviour.
Another notion called \emph{strong bisimulation} exists. It is
nevertheless too restrictive in our context because it imposes a
strict matching of the $\tau$ actions.  Also note the notion of
\emph{congruence}, an observational equivalence which should be
preferred when one wants services to be equivalent \emph{in any
  context}, {\it i.e.}, in all possible systems using them.

\subsection{Verifying Properties}

The properties of interest in concurrent systems typically involve
reasoning on the possible scenarii that the system can go through. An
established formalism for expressing such properties is given by
\emph{temporal logics}\footnote{This name should not give the
  impression that these logics introduce a quantitative notion of
  time, they are indeed used to express constraints on the possible
  executions of a system.} like
CTL$\star$~\cite{Manna-Pnueli:BOOK:1995}. These logics present constructs
allowing to state in a formal way that, for instance, all scenarii
will respect some property at every step, or that some particular
event will eventually happen, and so on.

An introduction to temporal logic goes beyond the aims of this paper, but it
suffices to say that a number of classical properties typically appear as
patterns in many applications. Reusing them diminishes the need to learn all
subtleties of a new formalism. The most noticeable properties are:

\smallskip
$\bullet$ \textbf{Safety properties}, which state that an undesirable situation
  will never arise. For instance, the requirements can forbid that the system
  reserves a room without having received the credit information from the
  bank;

$\bullet$ \textbf{Liveness properties}, which state that some actions will
  always be followed by some reactions; a typical example is to check that
  every request for a room will be acknowledged.

The techniques used to check whether a system respects temporal logic
properties are referred to as \emph{model checking} methods
\cite{Clarke-Grumberg-Peled:BOOK:2000}. Several tools exist and can be
used to model-check abstract descriptions of services, {\it e.g.},
CADP, or SPIN.

\section{Compatibility and Adaptation}
\label{section:compatibility}

\subsection{Compatibility Notions}
\label{section:compnotions}

Compatibility aims at ensuring that services will be able to interact
properly, that is satisfy a specific criterion on observable actions
and terminate in final states. Typically, compatibility is needed at
design-time as a previous step (discovery) in a service composition
construction in order to avoid erroneous executions at run-time.
Substitutability is a similar issue and aims at replacing one service
by another without introducing flaws. Substitutability can be checked
using equivalence-checking techniques presented in the previous
section.  Compatibility checking, if defined in a formal way, can be
automated using state space exploration tools such as CADP or SPIN, or
rewriting-based tools such as Maude.

In the rest of this subsection, we introduce three notions of
compatibility, namely deadlock-freeness,
unidirectional-complementarity and unspecified-receptions, that make
sense in the Web services area. These notions have often been studied
in the
literature~\cite{DZ-ACM83,YS-ACM97,CanalPT01,tes04,DOS-Foclasa09,OS-WCSI10}.

\begin{figure}
\centering
\includegraphics[width=0.6\linewidth,clip]{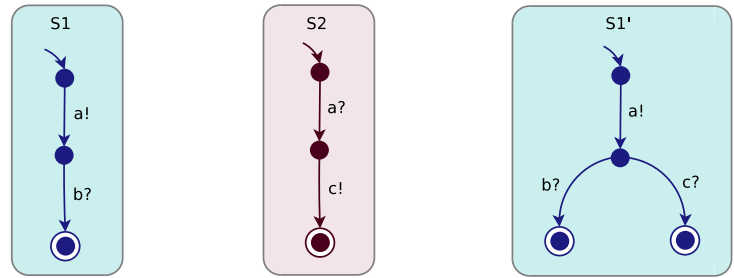}
\caption{Deadlock-freeness compatibility}
\label{fig:dead}
\end{figure}

\noindent\textbf{Deadlock-freeness.} This notion says that two service
protocols are compatible if and only if, starting from their initial
states, they can evolve together until reaching final states.
Figure~\ref{fig:dead} presents a simple example to illustrate this
notion. {\sf S1} and {\sf S2} are not compatible because after
interacting on action {\sf a}, both services are stuck. On the other
hand, {\sf S1'} and {\sf S2} are deadlock-free compatible since they
can interact successively on {\sf a} and {\sf c}, and then both
terminate into a final state.

\medskip

\noindent \textbf{Unidirectional-complementarity.}  Two services are
compatible with respect to this notion if and only if there is one
service which is able to receive (send, respectively) all messages
that its partner expects to send (receive, respectively) at all
reachable states. Hence, the ``bigger'' service may send and receive
more messages than the ``smaller'' one. Additionally, both services
must be free of deadlocks.
This notion is different to what is usually called simulation or
preorder relation~\cite{MilCC} because the two protocols under
analysis here aim at being composed, and accordingly present opposite
directions. However, both definitions share the inclusion concept: one
of the two protocols is supposed to accept all the actions that the
other can do.
Figure~\ref{fig:sim} first shows two services {\sf S1} and {\sf S2}
which respect this unidirectional-complementarity compatibility: all
actions possible in {\sf S1} can be captured by {\sf S2}. However,
{\sf S2} does not complement {\sf S1'} because {\sf S2} is not able to
synchronize on action {\sf c} with {\sf S1'}.

\medskip

\noindent \textbf{Unspecified-receptions.}  This definition requires
that if one service can send a message at a reachable state, then the
other service must receive that emission. Furthermore, one service is
able to receive messages that cannot be sent by the other service,
{\it i.e.}, there might be additional unmatched receptions.  It is
also possible that one protocol holds an emission that will not be
received by its partner as long as the state from which this emission
goes out is unreachable when protocols interact together.
Additionally, both services must be free of deadlocks.
In Figure~\ref{fig:unspecified}, {\sf S1} and {\sf S2} are not
compatible because {\sf S1} cannot receive all actions that {\sf S2}
can send ({\sf c!}). But {\sf S1'} and {\sf S2} are compatible because
all emissions on both sides have a matching reception on the other.

\medskip

The reader interested in the formal definitions for these
compatibility notions can refer to~\cite{tes04,DOS-Foclasa09}. 

\subsection{Compatibility Degree}
\label{section:compdegree}

Most of the approaches existing for checking compatibility return a
``True'' or ``False'' result to detect whether services are compatible
or not. Unfortunately, a Boolean answer is not very helpful for many
reasons.  First, in real world case studies, there will seldom be a
perfect match, and when service protocols are not compatible, it is
useful to differentiate between services that are slightly
incompatible and those that are totally incompatible. Furthermore, a
Boolean result does not give a detailed measure of which parts of
service protocols are compatible or not. To overcome the
aforementioned limits, a new solution aims at measuring the
compatibility degree (or similarity degree if the idea is to replace
and not to compose services) of service interfaces. This issue has
been addressed by a few recent works, see for
instance~\cite{SokolskyKL06,NejatiEtAl07,Lohmann08,Ait-Bachir-ICSOC08,wu2009computing,OSP10}.

\begin{figure}
      \centering
      \includegraphics[width=0.6\linewidth,clip]{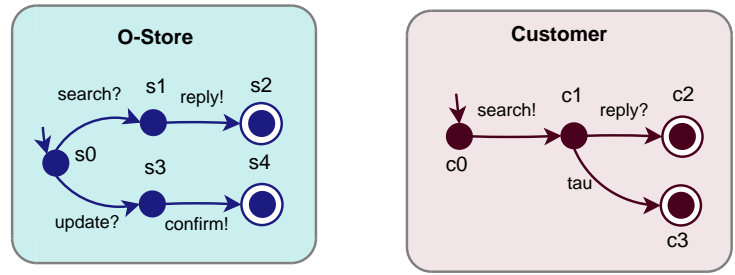}
      \caption{An online store}
      \label{fig:excd}
\end{figure}

Let us illustrate with a simple example (Fig.~\ref{fig:excd}) the kind
of results one can compute with these compatibility measuring
approaches. Here, we use the compatibility measuring algorithms
presented in~\cite{OSP10}. This approach takes as input two STSs and
computes a compatibility degree for each global state, {\it i.e.,}
each couple of states $(s_i, s_j)$ with $s_i \in S_1$ and $s_j \in
S_2$. All compatibility scores range between 0 and 1, where 1 means a
perfect compatibility.  To measure the compatibility of two service
protocols, the protocol compatibility degrees are computed for all
possible global states using a set of static compatibility measures.
This work uses three static compatibility measures, namely state
natures, labels, and exchanged parameters. These measures are used
next to analyse the behavioural part (ordering of labels) of both
protocols.  Intuitively, two states are compatible if their backward
and forward neighbouring states are compatible, where the backward and
forward neigbours of state $s'$ in transitions $(s,l,s')$ and
$(s',l',s'')$ are respectively the states $s$ and $s''$.  Hence, in
order to measure the compatibility degree of two service protocols, an
iterative approach is considered which propagates the compatibility
degree from one state to all its neighbours. This process is called
compatibility flooding.

Table~\ref{table:matrix} shows the matrix computed for the example
depicted in Figure~\ref{fig:excd} according to the
unidirectional-complementarity notion. Let us comment the
compatibility of states ${\sf c0}$ and ${\sf s0}$. The measure is
quite high because both states are initial and the emission {\sf
  search!} at ${\sf c0}$ perfectly matches the reception {\sf search?}
at ${\sf s0}$. However, the compatibility degree is less than 1 due to
the backward propagation of the deadlock from the global state $({\sf
  s1},{\sf c3})$ to $({\sf s1},{\sf c1})$, and then from $({\sf
  s1},{\sf c1})$ to $({\sf s0},{\sf c0})$.

\begin{table}
\center
\begin{tabular}{c|cccccc|}
			
     & {\sf s0} & {\sf s1} & {\sf s2} & {\sf s3} & {\sf s4}\\
\hline

{\sf c0}&	\textbf{0.78}& 0.01&	0.01&	0.01& 	0.01\\

{\sf c1}&	0.01	&0.68	&0.01	&0.35	&0.01\\

{\sf c2}&	0.01	&0.01&	0.90	&0.01	&0.67\\

{\sf c3}&0.01	&0.45&	0.76&	0.35	&0.76
\end{tabular}
\caption{The compatibility matrix computed for the example in Figure~\ref{fig:excd}}
\label{table:matrix}
\end{table}

\subsection{Service Adaptation}
\label{section:adaptation}

While searching a service satisfying some specific requirements, one
can find a candidate which exhibits the expected functionality but
whose interface does not exactly fit in the rest of the system.  {\it
  Software Adaptation}~\cite{BeckerDagstuhl2005} is a very promising
solution to compose in a non-intrusive way black-box components or
(Web) services although they present interface mismatches. Adaptation
techniques aim at automatically generating new components called {\it
  adaptors}, and usually rely on an {\it adaptation contract} which is
an abstract description of how mismatches can be worked out. All the
messages pass through the adaptor which acts as an orchestrator, and
makes the involved services work correctly together by compensating
mismatches. The generation of this adaptor is a complicated task,
especially when interfaces take into account a behavioural description
of the service execution flow. Recently, several approaches have been
proposed to generate service adaptors, see for
example~\cite{BrogiICSOC06,BenatallahWWW07,MateescuPS08,Canal-Poizat-Salaun-08,AalstMSW09}.

\begin{figure}[h]
  \centerline{\includegraphics[width=0.75\linewidth]{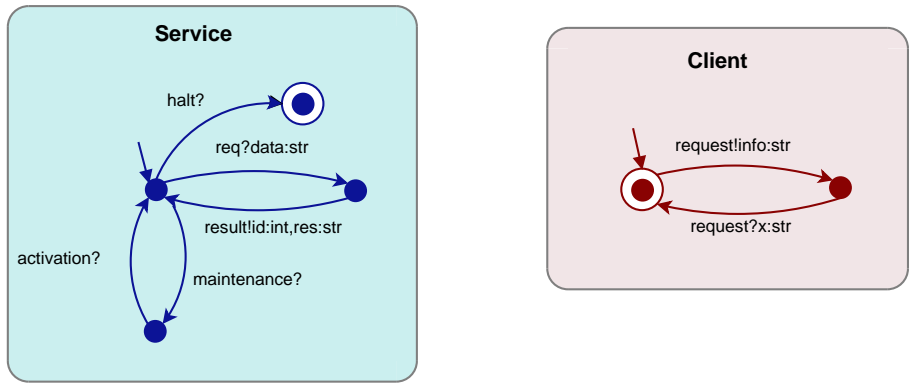}}
\caption{An SQL service}
\label{fig:interfaces}
\end{figure}

Figure~\ref{fig:interfaces} gives an example: the first interface
corresponds to an SQL service which can receive ({\sf req?})  and
answer ({\sf result!})  requests, stops ({\sf halt!}), or halts
temporarily for maintenance purposes ({\sf maintenance?}  and {\sf
  activation?}). The client can submit requests ({\sf request!}), and
receive responses ({\sf request?}).

Several notations exist for writing adaptation contracts. In this
paper, we use {\it vectors}~\cite{MateescuPS08} which specify
interactions between several services. They express correspondences
between messages, like bindings between ports, or connectors in
architectural descriptions.  Each label appearing in one vector is
executed by one service and the overall result corresponds to an
interaction between all the involved services.  Furthermore,
variables are used as placeholders in message parameters. The same
variable name appearing in different labels (possibly in different
vectors) enables one to relate sent and received arguments of
messages.

\begin{figure}[h]
  \centerline{\includegraphics[width=0.75\linewidth]{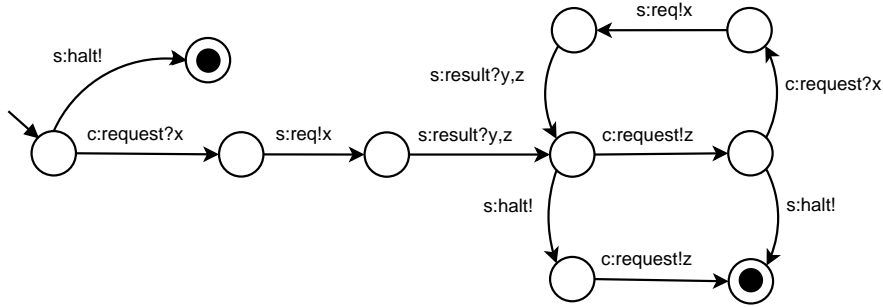}}
\caption{The adaptor protocol for the SQL example}
\label{fig:adaptor}
\end{figure}

As far as our example is concerned, the following vectors constitute a
contract from which the adaptor protocol given in
Figure~\ref{fig:adaptor} is automatically generated by using
techniques and tools presented in~\cite{MateescuPS08}.
This approach respectively generates (i) LOTOS code\footnote{LOTOS is
  a value passing process algebra proposed in the late 80s,
  see~\cite{BolognesiB87} for more details.} for service interfaces
and the contract, and (ii) the corresponding state space by applying
on-the-fly simplification (deadlock suppression) and reduction
techniques ($\tau$ transition removal).

\smallskip

{\sf V1}      = $\langle \mathrm{{\sf s}}\!:\! \mathrm{{\sf req?X}}; \mathrm{{\sf c}}\!:\! \mathrm{{\sf request!X}} \rangle \;\;\;\;\;\;\;\;\;\;$ 
{\sf V2}      = $\langle \mathrm{{\sf s}}\!:\! \mathrm{{\sf result!Y,Z}}; \mathrm{{\sf c}}\!:\! \mathrm{{\sf request?Z}} \rangle \;\;\;\;\;\;\;\;\;\;$ 
{\sf V3}      = $\langle \mathrm{{\sf s}}\!:\! \mathrm{{\sf halt?}} \rangle$

From adaptor protocols, either a central adaptor can be implemented,
or several service wrappers can be generated to distribute the
adaptation. In the former case, the implementation of executable
adaptors from adaptor protocols can be achieved for instance using
techniques presented in~\cite{MateescuPS08} and~\cite{CuboSCPP08} for
BPEL and Windows Workflow Foundation, respectively. In the latter
case, each wrapper constrains its service functionality to make it
respect the adaptation contract~\cite{salaun-SEFM-2008}.

\section{Realizability and Conformance}
\label{section:realizability}

Interactions among a set of services involved in a new system can be
described from a global point of view using {\it choreography}
specification languages. Several formalisms have already been proposed
to specify choreographies: WS-CDL, collaboration diagrams, process
calculi, BPMN, SRML, etc. Given a choreography specification, it would
be desirable if the local implementations, namely {\it peers}, can be
automatically generated via projection. However, generation of peers
that precisely implement the choreography specification is not always
possible: This problem is known as {\it realizability}. A related
problem is known as {\it conformance} where the question is to check
whether a choreography and a set of service implementations (not
obtained by projection from the choreography) produce the same
executions.

\begin{figure}[ht]
\centerline{\includegraphics[width=0.9\linewidth]{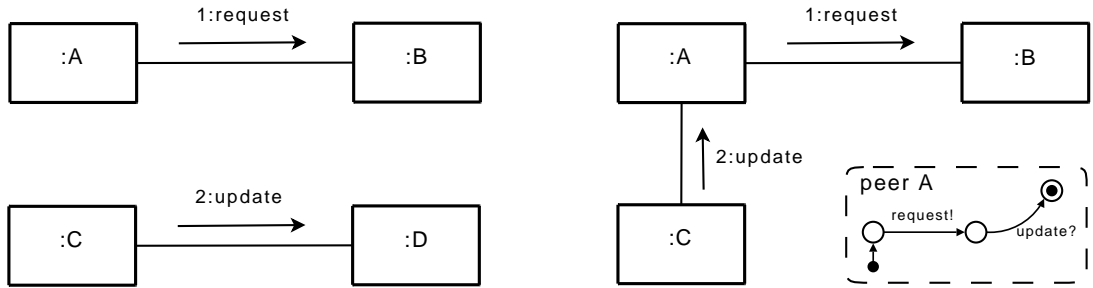}}
\caption{Examples of unrealizable collaboration diagrams}
\label{figure:unrealizable}
\end{figure}

A couple of unrealizable collaboration diagrams~\cite{Bultan-Fu-08}
are presented in Figure~\ref{figure:unrealizable}. The first one (left
hand side) is unrealizable because it is impossible for the peer {\tt
  C} to know when the peer {\tt A} sends its {\tt request} message
since there is no interaction between {\tt A} and {\tt C}. Hence, the
peers cannot respect the execution order of messages as specified in
the collaboration diagram. The second one is slightly more subtle
because this diagram is realizable for synchronous communication, and
unrealizable for asynchronous communication. Indeed, in case of
synchronous communication, the peer {\tt C} can synchronize
(rendez-vous) with the peer {\tt A} only after the {\tt request}
message is sent, so the message order is respected.  This is not the
case for asynchronous communication since {\tt A} cannot block {\tt C}
from sending the {\tt update} message.  Hence, {\tt C} has to send the
{\tt update} message to {\tt A} without knowing if {\tt A} has sent
the {\tt request} message or not.  Therefore, the correct order
between the two messages cannot be satisfied. We also show in
Figure~\ref{figure:unrealizable} the LTS generated for peer {\tt A} by
projection.

Several works aimed at studying and defining the realizability (and
conformance) problem for choreography, here are a few
references~\cite{KP-FORTE06,Busi-Coordination2006,LiEtAlTASE07,Fu05,Bultan-Fu-08}.
In~\cite{Busi-Coordination2006,LiEtAlTASE07}, the authors define
models for choreography and orchestration, and formalize a conformance
relation between both models.  Other
works~\cite{Carbone-ESOP07,ZongyanWWW07} propose well-formedness rules
to enforce the specification to be realizable. A few
works~\cite{ZongyanWWW07,SB-IFM09} also propose to add messages in
order to implement unrealizable choreographies.  Fu {\it et
  al.}~\cite{FuBS04} proposed three conditions (lossless join,
synchronous compatible, autonomous) that guarantee a realizable
conversation protocol under asynchronous communication. These
conditions have been implemented in the WSAT tool~\cite{FuBS-CAV04}
which takes a conversation protocol as input, and says if it satisfies
the three realizability conditions. \cite{SuBFZ07} discusses some
interesting open issues in this area.

\section{Concluding Remarks}
\label{section:conclusion}

In this paper, we have surveyed some issues in Web services which
require analysis and verification techniques. Using these techniques
seems natural when one wants to ensure that a composition of services
will work correctly or satisfy some high-level requirements. But they
have also other applications in SOC, {\it e.g.}, to check the
compatibility of a service with a possible client (discovery), or to
generate some service adaptors if some interface mismatches prevent
their direct composition. Last, we have showed that when specifying a
system using choreography languages, some analysis are useful to check
that the corresponding distributed implementation will behave as
described in the global specification.

We would like to conclude with a few challenges which are still some
open issues, as far as analysis techniques are concerned, in the Web
services domain. All these challenges assume an asynchronous
communication model (that is based on message queues).  A few works
already exist, in~\cite{Fu05} for example the authors define a
synchronizability condition which makes systems under asynchronous
communication verifyable with tools working with synchronous
comunication. Some sufficient conditions have also been proposed to
guarantee the realizability of conversation protocols~\cite{FuBS04}.
Nevertheless, in both works, if these conditions are not satisfied,
nothing can be concluded on the system being analysed.

Some open challenges assuming an asynchronous communication model are
the following: (i)~providing automated techniques to check the
compatibility of two or more services, (ii)~checking the adaptability
of a set of services being given an adaptation contract, and if the
system is adaptable, generating the corresponding adaptor,
(iii)~finding a decidable algorithm for checking the realizability of
a choreography specification language with loops (such as conversation
protocols~\cite{FuBS04}).

\medskip

{\bf Acknowledgements.} The author would like to thank Meriem Ouederni
for her comments on a former version of this paper.

\bibliographystyle{eptcs}
\bibliography{biblio}

\end{document}